\documentclass[12pt]{article}
\date{}
\usepackage{amsmath}
\usepackage{amsfonts}
\usepackage{amssymb}
\usepackage{graphicx}

\newcommand{\bea}{\begin{eqnarray}}
\newcommand{\eea}{\end{eqnarray}}
\newcommand{\be}{\begin{equation}}
\newcommand{\ee}{\end{equation}}

\renewcommand{\b}{\beta}

\newcommand{\dsl}{\pa \kern-0.5em /}

\newcommand{\pa}{\partial}

\newcommand{\nn}{\nonumber\\}

\def\be{\begin{equation}}
\def\ee{\end{equation}}

\begin{document}

\title{{\bf{Holographic complexity for Lifshitz system}}}

\author{
{\bf {\normalsize Sourav Karar}$^{a,b}$\thanks{sourav.karar91@gmail.com}},
{\bf {\normalsize Sunandan Gangopadhyay}$^{c}$\thanks{sunandan.gangopadhyay@gmail.com, sunandan@associates.iucaa.in }}\\
$^{a}$ {\normalsize Department of Physics, Government General Degree College, Muragachha 741154, Nadia, India}\\
$^{b}${\normalsize S. N. Bose National Centre for Basic Sciences, Kolkata 700106, India}\\
$^{c}$ {\normalsize Department of Physical Sciences, Indian Institute of Science Education $\&$ Research,}\\
{\normalsize Kolkata, Mohanpur 741246, Nadia, India}\\
\\[0.3cm]}

\maketitle
\begin{abstract}
\noindent The holographic complexity of a $3+1$-dimensional Lifshitz spacetime having a scaling symmetry is computed.
The change in the holographic complexity between the excited state and the ground state is then obtained.
This is then related to the changes in the energy and the entanglement chemical potential of the system.
The calculation is carried out for both the values of the dynamical scaling exponent $z$ in the Lifshitz spacetime.
The relations has a very similar form to the corresponding relation involving the change in entanglement entropy
known to be an analogous relation to the first law of thermodynamics.

\end{abstract}

\newpage

\section{Introduction}  

Quantum entanglement \cite{Bombelli:1986rw}-\cite{Eisert:2008ur} between two subsystems $A$ and  $B$ for a given pure state can be characterized 
by a quantity called entanglement entropy (EE) and has been an important topic of intense research in the field
of quantum information. It has also been realized that quantum information may be an useful tool to
study black hole physics. The AdS/CFT correspondence on the other hand has been one of the most 
remarkable theoretical developments in providing deep insights in the realm of quantum gravity, black hole physics
and strongly coupled condensed matter systems. This has led to the idea of constructing holographic gravitational 
descriptions of quantities relevant in the field of quantum information.

A holographic description of quantum entanglement known as holographic EE \cite{Ryu:2006bv, Ryu:2006ef} has been an important line 
of recent activity and has proved to be useful in computing the EE of conformal field theories \cite{Calabrese:2004eu}-\cite{Calabrese:2009qy}. 
The usefulness of EE is also manifest in describing systems away from equilibrium.
 In this context, an important question that has been asked in the literature \cite{Bhattacharya:2012mi}
 is whether an analogous relation to the first law of thermodynamics holds good in this case also. The answer to this question was
 found in the affirmative \cite{Bhattacharya:2012mi}. The change in EE was found to be proportional to the change in the energy of the system 
 for sufficiently small subsystems. The constant of proportionality was found to be related to the size of the entangling region
 and was identified with the inverse of some temperature, named to be the entanglement temperature.
 The resemblance of the change in EE with the first law of thermodynamics was also confirmed in \cite{Allahbakhshi:2013rda} where
 they computed the change in EE between an excited state of an AdS spacetime and pure AdS spacetime. However, the analysis
 in these cases were restricted to the relativistic systems. In \cite{skarar}, the question was addressed in the case of a non-relativistic system, the
 well known Lifshitz system in $3+1$- dimensions. The excited state that was considered there was non-isotropic.
 This resulted in the addition of an extra term in the change in entropy and was identified with an entanglement chemical potential.
 
 Quantum complexity is another quantity which has been proved its importance in understanding the properties of horizons
 of black holes. The quantity intuitively indicate how difficult it is to map a system in a given state to another state.
 Motivated from studies in the holographic description of EE, investigations have been initiated in providing a 
 holographic dual description of quantum complexity using the AdS/CFT correspondence. A prescription
 to compute the holographic complexity (HC) was proposed in \cite{sus1, sus2}. The prescription tells that for subsystem
  $A$ in the boundary, if $V(\gamma)$  denotes the volume enclosed by the minimal hyper surface in the bulk, 
  then the holographic  complexity is given by 
  \be
  \mathcal{C}_A =\frac{V(\gamma)}{8\pi R G}
  \ee
where  $R$ is the radius of curvature of the spacetime. In case of an asymptotically AdS spacetime, $R$
would be the AdS radius. This quantity was computed in $AdS_{d+2}$ geometry in \cite{Mohsen}.

In this paper, we shall compute the HC for the non-relativistic Lifshitz system in $3+1$- dimensions.
We shall first compute it for the pure Lifshitz spacetime and then for the excited state. This would give
us the change in the HC between the perturbed and the pure Lifshitz spacetimes. We would then try to relate this change 
with changes in the energy and entanglement chemical potential. We shall carry out this analysis for both the
values of the dynamical scaling exponent $z$ appearing in the Lifshitz spacetime.

The paper is organized as follows. In the next section, we compute the change in the HC between 
an excited state of the Lifshitz spacetime and the ground state. We then relate it to the components of
the holographic stress tensor. We conclude in section 3.

\vspace{.5cm}


\section{Holographic complexity for Lifshitz spacetime} 

In this section we will first  
compute the HC, that is the Ryu-Takayanagi (RT) volume with a strip geometry for the unperturbed Lifshitz spacetime. 
We shall then compute the HC for the perturbed Lifshitz spacetime which corresponds to the excited state. 
This is in turn would lead to the change in the HC between the excited and the unperturbed Lifshitz spacetimes. 
To begin with we shall present a brief description of the set up needed to compute the holographic complexity for the Lifshitz spacetime.
The four dimensional Lifshitz metric reads \cite{Ross:2009ar}
\bea \label{lmet}
ds^2 &=& - r^{2z} dt^2 + r^2 (dx^2 + dy^2) + \frac{dr^2}{r^2}\nn
 \quad A &=& \alpha r^z dt ,\quad \alpha^2 = \frac{2(z-1)}{z}~.
\eea
The spacetime is the gravitational dual of a $2+1$-dimensional quantum many body system with a Lifshitz symmetry near its quantum critical point.
This solution can be obtained from the equations of motion following from the action \cite{Kachru:2008yh, Taylor:2008tg}
\be\label{action1}
S = \frac{1}{16\pi G_4}\int d^4x \sqrt{-g}\left(R - 2\Lambda - \frac{1}{4}F_{\mu\nu}F^{\mu\nu}
- \frac{1}{2} m^2 A_\mu A^\mu\right)
\ee
with the choice $\Lambda = -\frac{1}{2}(z^2+z+4)$ for the cosmological constant and $m^2 = 2z$, $A_\mu$ being a massive gauge field.
\noindent It is evident that the above metric has the 
scaling symmetry $t \to \lambda^z t,\, x \to \lambda x,\, y \to \lambda y,\,  r \to \lambda^{-1} r$, where $z$ is the dynamical scaling exponent. 

\noindent The entangling region in the boundary is taken 
to be  a straight belt with width $\ell$ such that $-\frac{\ell}{2} \leq x \leq \frac{\ell}{2}$ and $0 \leq y \leq L$, where $L$ is the extent of the subsystem in the other spatial direction. 
Since the strip has translational invariance along the direction $y$, one can describe the profile of the extremal surface by
$x=x(r)$. With this set up in place, we can now proceed to compute the RT volume enclosed by the minimal surface extending from the boundary into the bulk.
In this case, this is given by
\bea\label{volume1}
V^{(0)} &=& 2L\int_{r^{(0)}_t}^{\infty} dr\; r\; x(r)
\eea
where $r^{(0)}_t$ is the turning point at which $r'(x)=0$.
To obtain the minimal surface $x(r)$, we write down the RT area functional considering $r=r(x)$. This is given by \cite{Ryu:2006bv,Ryu:2006ef}
\begin{eqnarray}
\label{are}
A^{(0)}&=& \int_{-\frac{\ell}{2}}^{\frac{\ell}{2}} dx \int_{0}^{L}  dy \sqrt{r'(x)^{2} + r^4}\nonumber\\
&=& 2 L\int_{0}^{\ell/2}dx~r^2\sqrt{1+\frac{r'(x)^2}{r^4}}
\end{eqnarray}
where $\prime$ denotes derivative with respect to $x$.
The minimization of this area functional determines the function
$r'(x)$ which reads
\begin{eqnarray}
\label{area1}
r'(x)=\frac{r^4}{r^{(0)2}_t}\sqrt{1-\frac{r^{{(0)}4}_t}{r^4}}~.
\end{eqnarray}
 This now leads to
\begin{eqnarray}
\label{area2}
x(r)=\int_{r^{(0)}_{t}}^{r}du~\frac{r^{(0)2}_{t}}{u^{4}}\frac{1}{\sqrt{1-\frac{r^{(0)4}_{t}}{u^4}}}~.
\end{eqnarray}
Substituting the above expression for $x(r)$ in eq.(\ref{volume1})
and putting a cut-off $\delta$ for the $r$ integral, we have
\bea\label{volume2}
V^{(0)} &=& 2L\int_{r^{(0)}_t}^{\delta} dr\; r\;\int_{r^{(0)}_{t}}^{r}du~\frac{r^{(0)2}_{t}}{u^{4}}\frac{1}{\sqrt{1-\frac{r^{(0)4}_{t}}{u^4}}}\nn
&=& \frac{\sqrt{\pi} \; \Gamma \left (\frac{3}{4}\right) \; L \; \delta^2}{\Gamma\left( \frac{1}{4} \right) \; r^{(0)}_t} 
-\frac{\sqrt{\pi} \; \Gamma \left(\frac{5}{4} \right) \; L \; r^{(0)}_t}{\Gamma\left(\frac{3}{4} \right)} 
\eea
where we have ignored terms of order $1/\delta$ since they are small.

\noindent Substituting eq.(\ref{area1}) in eq.(\ref{are})
and once again using the cut-off $\delta$ for the $r$ integral and ignoring terms of order $1/\delta$, we get
\begin{eqnarray}
\label{marea}
A^{(0)}&=& 2L \int_{r^{(0)}_t}^{\delta} dr \frac{\left(\frac{r}{r^{(0)}_t}\right)^2}{\sqrt{\left(\frac{r}{r^{(0)}_t}\right)^4 -1}}\nn
&=& 2 L \delta -\frac{5}{3} L r^{(0)}_{t} ~.
\end{eqnarray}
From eq.(\ref{area1}), we also have
\be\label{ell}
\frac{\ell}{2} = \int_{0}^{\frac{\ell}{2}} dx =\int_{r^{(0)}_t}^{\infty} dr \frac{1}{r^2 \sqrt{\left(\frac{r}{r^{(0)}_t}\right)^4 -1}}
=\frac{\sqrt{\pi} \; \Gamma\left(\frac{3}{4}\right)}{\Gamma\left(\frac{1}{4}\right) r^{(0)}_t}~.
\ee
We know that the HC is proportional to the volume enclosed by the minimum surface extending from the boundary into the bulk,
 which in this case is given by
\begin{eqnarray}
\mathcal{C}^{(0)}_A &=&\frac{V^{(0)}}{8\pi G_4}\nn
&=&\frac{\sqrt{\pi}}{8\pi G_4}\left[
 \frac{\; \Gamma \left (\frac{3}{4}\right) \; L \; \delta^2}{\Gamma\left( \frac{1}{4} \right) \; r^{(0)}_t} 
-\frac{\; \Gamma \left(\frac{5}{4} \right) \; L \; r^{(0)}_t}{\Gamma\left(\frac{3}{4} \right)}\right].                            
\end{eqnarray}
This is the HC for the ground state of the Lifshitz spacetime.

\noindent We shall now proceed to calculate the HC for the perturbed Lifshitz metric \cite{Ross:2009ar} 
\bea \label{lmetper}
ds^2 &=& - r^{2z} (1+h_{tt}(r))dt^2 + r^2 (1+h_{xx}(r))dx^2 + r^2(1+h_{yy}(r))dy^2 + \frac{dr^2}{r^2}  \nn 
& & + 2 [-r^{2z} v_{1x}(r) + r^2 v_{2x}(r)] \; dt dx +2 [-r^{2z} v_{1y}(r) + r^2 v_{2y}(r)] \; dt dy + 2 r^2 h_{xy}(r)\; dx dy\nn
A &=& \alpha r^z [(1+ a_t(r) + \frac{1}{2}h_{tt}(r))dt + v_{1x}(r) dx + v_{1y}(r) dy]. 
\eea 
This can be considered as the excited state of the Lifshitz spacetime. 
An asymptotically Lifshitz spacetime results when 
$h_{tt}(r)$, $h_{xx}(r)$, $h_{yy}(r)$, $v_{1x}(r)$, $v_{2x}(r)$, $v_{1y}(r)$, $v_{2y}(r)$, $h_{xy}(r)$, and $a_t(r)$ $\rightarrow 0$ as $r\to \infty$.
One can now define
  \bea\label{defn}
  h_{tt}(r) &=& f(r)\nn
  h_{xx}(r) &=& k(r) + {t_d}(r) \nn
  h_{yy} (r) &=& k(r) - {t_d}(r) \nn
  a_t &=& j(r).
  \eea
Substituting this in eq.(\ref{action1}), the linearized action 
can be obtained \cite{Ross:2009ar}. 
The solutions obtained by solving the equations of motion (in the radial gauge)
resulting from the linearized action have the form \cite{Ross:2009ar, Ross:2011gu, Andrade:2013wsa} for $z=2$ 
\bea\label{z2}
 j(r) &=& -\frac{c_1 + c_2 \ln r}{r^4}, \nn
 f(r) &=& \frac{4 c_1 - 5 c_2 + 4 c_2 \ln r}{12 r^4}, \nn
 k(r) &=& \frac{4 c_1 + 5 c_2 + 4 c_2 \ln r}{24 r^4}, \nn
t_d(r) &=& \frac{t_{d2}}{r^4}
\eea
and for $z\neq 2$
\bea\label{zneq2}
 j(r) &=& -\frac{(z+1)c_1}{(z-1)r^{z+2}} -
 \frac{(z+1)c_2}{(z-1)r^{\frac{1}{2}(z+2+\beta_z)}}, \nn
 f(r) &=& 4\frac{1}{(z+2)} \frac{c_1}{r^{z+2}} + 2
 \frac{(5z-2-\beta_z)}{(z+2+\beta_z)}
 \frac{c_2}{r^{\frac{1}{2}(z+2+\beta_z)}}, \nn
 k(r) &=& 2 \frac{1}{(z+2)} \frac{c_1}{r^{z+2}} - 2
 \frac{(3z-4-\beta_z)}{(z+2+\beta_z)}
 \frac{c_2}{r^{\frac{1}{2}(z+2+\beta_z)}},\nn
t_d(r) &=& \frac{t_{d2}}{r^{z+2}}  
 \eea
where $\beta_z^2 = 9z^2-20z+20 = (z+2)^2 +8(z-1)(z-2)$,
$c_1$, $c_2$ and $t_{d2}$ are constants of integration.

\noindent With this set up in place, we can now move on to compute the HC for the perturbed Lifshitz metric (\ref{lmetper}).
This in turn would lead to the change in complexity due to the change
in the metric. In the computation of the complexity of the excited state, we shall keep the length of the entangling region $\ell$ fixed. Now as the entangling region is same but the metric is perturbed, hence 
the turning point changes. Let $r_t$ be the new turning point such that
$r_t =r^{(0)}_t +\delta r_t$, where $\delta r_t$ is the change in the turning point. 

\noindent As in the unperturbed case, we start by considering the minimal surface to be parametrized by $x=x(r)$. The volume in the part of the bulk geometry enclosed by the minimal surface is given by
 \be\label{volpert1}
 V= \int_{-\frac{\ell}{2}}^{\frac{\ell}{2}} dx \int_{0}^{L} dy 
 \int_{ r_t}^{\delta} dr \;
 	r \;x(r) \; \sqrt{1+h_{xx}(r) + h_{yy}(r)}.
 \ee
To determine $x(r)$, we once again start by writing down the area as a functional of the minimal surface $r=r(x)$. This reads
\begin{eqnarray}
A=2L\int_{0}^{\ell/2}dx \sqrt{r'^{2}(x)[1+h_{yy}(r)]+r^4 [1+h_{xx}(r)+h_{yy}(r)]}.
\label{n1}
\end{eqnarray}
Regarding $x$ as a time, one can easily obtain the Hamiltonian which does not depend on $x$. This in turn leads to
\begin{eqnarray}
r'(x)=\frac{\sqrt{\frac{r^8}{Q}[1+h_{xx}(r)+h_{yy}(r)]^2-r^4 [1+h_{xx}(r)+h_{yy}(r)]}}{\sqrt{1+h_{yy}(r)}}
\label{n2}
\end{eqnarray}
where $Q$ is a constant of integration. This gets fixed by the fact that $r'(x)=0$ at the turning point $r=r_{t}$. Hence we have
\begin{eqnarray}
Q=r_{t}^{4}[1+h_{xx}(r_t)+h_{yy}(r_t)].
\label{n3}
\end{eqnarray}
Eq.(\ref{n2}) determines the minimal surface $r=r(x)$ or $x=x(r)$. 

\noindent Integrating eq.(\ref{n2}) from $x=0$ to $x=\ell/2$ assuming that the $h$'s are small, we obtain upto first order in $h$
\begin{eqnarray}
\frac{\ell}{2}&=&\int_{r_t}^{\infty}\frac{dr}{r^2 f(r, r_t)}
\left\{1-h_{xx}(r) -\frac{1}{2}h_{yy}(r)+\frac{1}{2}[h_{xx}(r_t)+h_{yy}(r_t)]\right.\nonumber\\
&&\left. +\frac{[h_{xx}(r_t)+h_{yy}(r_t)-h_{xx}(r)-h_{yy}(r)]}{2f^{2}(r, r_t)}\right\}
\label{n4}
\end{eqnarray}
where $f^2 (r, r_t)=(r/r_t)^4 -1$. Now since we keep $\ell$ fixed, therefore we also have
\begin{eqnarray}
\frac{\ell}{2}=\int_{r^{(0)}_t}^{\infty}\frac{dr}{r^2 f(r, r^{(0)}_t)}~.
\label{n5}
\end{eqnarray}
Equating the above two expressions gives
\begin{eqnarray}
\delta r_t &=&-N r^{(0)}_{t}r_{t}\int_{r_t}^{\infty}\frac{dr}{r^2 f(r, r_t)}
\left\{h_{xx}(r)+\frac{1}{2}h_{yy}(r)-\frac{1}{2}[h_{xx}(r_t)+h_{yy}(r_t)]\right.\nonumber\\
&&\left. -\frac{[h_{xx}(r_t)+h_{yy}(r_t)-h_{xx}(r)-h_{yy}(r)]}{2f^{2}(r, r_t)}\right\}
\label{n6}
\end{eqnarray}
where
\begin{eqnarray}
N=\left\{\int_{1}^{\infty}\frac{d\xi}{\xi^2 \sqrt{\xi^4-1}}\right\}^{-1}=\frac{\Gamma(1/4)}{\sqrt{\pi}\Gamma(3/4)}~.
\label{n7}
\end{eqnarray}
Requiring $\delta r_t =0$ \cite{Allahbakhshi:2013rda} leads to the following condition on the perturbations
\begin{eqnarray}
\frac{[h_{xx}(r_t)+h_{yy}(r_t)]}{2f^{2}(r, r_t)}=
\frac{1}{[1+f^2 (r, r_t)]}\left\{h_{xx}(r)\left(1+\frac{1}{2 f^2 (r, r_t)}\right)+\frac{h_{yy}(r)}{2}\left(1+\frac{1}{f^2 (r, r_t)}\right)\right\}.
\label{n8}
\end{eqnarray}
Further since $\delta r_t=0$, we have $r_{t}=r^{(0)}_{t}$ in all the equations.
Using this condition in eq.(s)(\ref{n1}, \ref{n2}) yields
\begin{eqnarray}
A=A^{(0)}+L\int_{r^{(0)}_{t}}^{\delta}dr~\frac{[h_{yy}(r)+(\frac{r^{(0)}_t}{r})^4 h_{xx}(r)]}{\sqrt{1-(\frac{r^{(0)}_t}{r})^{4}}}
\label{n9}
\end{eqnarray}
and the same expression for $x(r)$ as in eq.(\ref{area2}).
Substituting the form of $x(r)$ (in eq.(\ref{area2})) in eq.(\ref{volpert1})
and keeping terms upto first order in $h$, we obtain
\be
\label{n10}
V = V^{(0)} +L \int_{r^{(0)}_t}^{\delta} dr\; r\; \; [h_{xx}(r) + h_{yy}(r)]
\int_{r^{(0)}_{t}}^{r}du~\frac{r^{(0)2}_{t}}{u^{4}}\frac{1}{\sqrt{1-\frac{r^{(0)4}_{t}}{u^4}}}~.
\ee
Note that since we are considering small perturbations around the background spacetime, $h(r)$'s are small. This in turn implies that the turning point $r_t$ is close to the boundary. However, since $r=\infty$ at the boundary and $\ell$ is inversely related to $r_t$, hence $\ell$ is small.

\noindent The change in HC due to the perturbation of the Lifshitz spacetime is given by
\begin{eqnarray}
\label{delc}
\Delta \mathcal{C}_A &=&\frac{\Delta V}{8\pi G_4}\nonumber\\
&=& \frac{L}{8\pi G_4}\int_{r^{(0)}_t}^{\delta} dr\; r\; \; [h_{xx}(r) + h_{yy}(r)]
\int_{r^{(0)}_{t}}^{r}du~\frac{r^{(0)2}_{t}}{u^{4}}\frac{1}{\sqrt{1-\frac{r^{(0)4}_{t}}{u^4}}}~.
\end{eqnarray}
Now we make use of the expressions for the perturbations to compute the HC explicitly.
For $z=2$, substituting the expressions 
for $h_{xx}(r)$ and $h_{yy}(r)$ from eq.(s)(\eqref{defn}) 
and \eqref{z2}), we obtain
\be\label{delcz2}
\Delta \mathcal{C}_A = \frac{\sqrt{\pi}\; \Gamma{ \left(\frac{5}{4}\right)} L}{ \; 8\pi G_4 \Gamma \left(\frac{7}{4}\right)r^{(0)3}_t}
	\left( \frac{c_1 +c_2 \ln{r^{(0)}_t}}{24} + \frac{3\pi +13}{288}c_2\right).
\ee
Our next step is to express the change in HC ($\Delta \mathcal{C}_A $) in terms of the holographic energy-momentum tensor.
The various components of the stress tensor can be found by varying the action with respect to the boundary metric. 
The expressions for the energy and pressure densities in terms of the functions defined earlier read \cite{Ross:2009ar}
\bea\label{energypressure1}
T_{tt} &=& -r^{z+2}\left[2r\partial_r k(r) + \alpha^2\left(zj(r) + r\partial_rj(r) + \frac{1}{2} r \partial_r f(r)\right)\right]\nn
T_{xx} &=& -2r^{z+2}\left[(z-1)j(r) - \frac{r}{2}\partial_r f(r) - \frac{r}{2} \partial_r k(r) - \frac{1}{2} (z+2) t_d(r)\right]\nn
T_{yy} &=& -2r^{z+2}\left[(z-1)j(r) - \frac{r}{2}\partial_r f(r) - \frac{r}{2} \partial_r k(r) + \frac{1}{2}(z+2) t_d(r)\right].
\eea
Substituting the forms of the functions for $z=2$ (given in eq.(\ref{z2}))
in the above expressions \cite{skarar}, we have
\bea\label{emz21}
\langle T_{xx} \rangle &=& \frac{1}{16 \pi G_4} \Big({\frac{4 c_2}{3}+4 t_{d2}} \Big)\\
\label{emz210}
\langle T_{tt} \rangle &=& \frac{1}{16 \pi G_4} {\frac{4 c_2}{3}}~.
\eea
Substituting $c_2$ from eq.(\ref{emz210}) in eq.(\ref{delcz2}), we obtain
\be\label{delcz20}
\Delta \mathcal{C}_A = \frac{\sqrt{\pi}\; \Gamma{ \left(\frac{5}{4}\right)} L}{ \; 192~ \Gamma \left(\frac{7}{4}\right)r^{(0)3}_t}
	\left( \frac{c_1 +c_2 \ln{r^{(0)}_t}}{\pi G_4} + (3\pi +13)\langle T_{tt} \rangle\right).
\ee
Now using the fact that the total energy, entanglement pressure, entanglement chemical potential of the excited state and charge are given by \cite{skarar}
\be\label{e1}
\Delta E = \int_{0}^{L}dy\int_{-\ell/2}^{\ell/2}dx~\langle T_{tt} \rangle  = L \ell \langle T_{tt} \rangle
\ee
\be\label{e2}
\Delta P_{x} = \langle T_{xx} \rangle
\ee
\be\label{e3}
\Delta \mu = \frac{1}{12\pi G_{4}}\left (c_1 + c_2\ln{r^{(0)}_t} \right)
\ee
\be\label{e4}
Q= m^2 \alpha L \ell=4 L \ell~;~ for ~ z=2
\ee
the above expression for the change in the HC can be recast as
\be\label{delcompz2}
\Delta \mathcal{C}_A = \frac{1}{T_{\rm ent}}\left [ B_1 \Delta \mu Q + B_2 \Delta E \right]
\ee
with  
\be\label{consb}
B_1= \frac{25 \;\Gamma \left(\frac{5}{4}\right) \Gamma \left(\frac{1}{4}\right)}
		{32 \pi \left(54-5\pi\right) \Gamma \left(\frac{7}{4}\right) \Gamma \left(\frac{3}{4}\right)}~,~
		\qquad
B_2 =\left( \frac{3\pi +13}{3}\right) B_1
\ee
and $T_{ent}$ is the entanglement temperature 
\be\label{tempz2}
T_{ent} = \frac{24 r_t^2}{\pi}\frac{25}{(324-30\pi)} = \frac{96 \Gamma^2\left(\frac{3}{4}\right)}{\ell^2 \Gamma^2\left(\frac{1}{4}\right)}
		\frac{25}{(324-30\pi)}
\ee
appearing in the expression for the first law of entanglement thermodynamics for an excited state of the Lifshitz spacetime \cite{skarar}
\be\label{thermoz2}
\Delta E = T_{\rm ent} \Delta S + \frac{10}{(54-5\pi)}\Delta P_x V - \frac{5}{(54-5\pi)}\Delta \mu Q.
\ee
It is to be noted that the change in pressure $\Delta P_x$ does not appear in the expression for the change in HC in contrast to the expression for the change in the holographic EE. Note that the origin of $\Delta P_x$ in the change in EE ($\Delta S$) is due to the presence of the
$t_{d2}$ term (related to $\Delta P_x$) in $\Delta S$ \cite{skarar} which arises from the combination 
$h_{yy}(r) + \left(\frac{r_t}{r}\right )^4 h_{xx}(r)$. However in the case of $\Delta \mathcal{C}_A$, the
combination $h_{xx}(r) + h_{yy}(r)$ arises which results in the cancellation of the $t_{d2}$ term responsible
for the origin of $\Delta P_{x}$.

\noindent We can recast eq.(\ref{delcompz2}) in another form.
Rewriting eq.\eqref{delcompz2} as 
\be\label{dez2}
\Delta E =\frac{1}{B_1} T_{ent}\Delta \mathcal{C}_A - \frac{B_2}{B_1} \Delta \mu Q
\ee
and equating this with eq.\eqref{thermoz2}, we have
\be\label{entcomz2}
\Delta \mathcal{C}_A = B_1 \Delta S + \frac{10\; B_1}{54-5\pi} \frac{\Delta P_x V}{T_{ent}} 
				    +\left( B_2 -\frac{5 B_1 }{54-5\pi} \right)\frac{\Delta \mu Q}{T_{ent}}~.
\ee
This expression relates the change in the HC with the change in holographic EE, change in pressure and the change in entanglement chemical potential.

\noindent We now compute the change in the HC from eq.\eqref{delc} for $z\neq2$. For this we substitute the expressions for $h_{xx}(r)$ and $h_{yy}(r)$ from eq.(s)\eqref{defn} and \eqref{zneq2} in eq.\eqref{delc}. The change is
\be\label{delczneq2}
\Delta \mathcal{C}_A = \frac{\sqrt{\pi} L}{8 \pi G_4 \; r_t^{(0)z+1}}
				   \left[ \frac{c_1 \Gamma \left(\frac{z+3}{4}\right)}{z(z+2) \Gamma \left(\frac{z+5}{4}\right)}
				  + \frac{2 c_2 (4+\beta_z -3z)\Gamma \left(\frac{z+\beta_z +4}{8}\right)}
				   {(z+\beta_z +2)(z+\beta_z-2) \Gamma \left(\frac{z+\beta_z +8}{8}\right)}\right ].
\ee	
The energy-momentum tensor and the entanglement chemical potential for $z\neq2$ can be obtained from eq.\eqref{energypressure1} using eq.\eqref{zneq2} and they read
\bea\label{emzneq2}
\langle T_{tt} \rangle  &=& \frac{1}{4\pi G_4}\left(\frac{z-2}{z}\right) c_1 \\
\label{emzneq2r}
\Delta \mu &=&  \frac{1}{16\pi G_4}\frac{\alpha}{[z-1]}\left[4c_1 + c_2 z(4+\beta_z - 3z) 
r_t^{(0)\frac{1}{2}(z+2-\beta_z)}\right].
\eea
We now invert the above expressions to obtain $c_1$ and $c_2$ as functions of
$\langle T_{tt} \rangle$ and $\Delta \mu$. These read
\bea\label{emzneq2inv}
c_1 &=& 4\pi G_4\frac{ \;z}{z-2} \langle T_{tt} \rangle \nn
c_2 &=& \frac{16\pi G_4\left[(z-1)(z-2) \Delta \mu - \alpha z \langle T_{tt} \rangle\right]}{\alpha z(z-2)(4+\beta_z-3z)r_t^{\frac{1}{2}(z+2-\beta_z)}}~.
\eea
Now the entanglement temperature appearing in the first law of entanglement thermodynamics for Lifshitz system with $z\neq2$ reads \cite{skarar}
\be\label{tempzneq2}
T_{\rm ent} = \frac{2 \Gamma\left(\frac{3}{4}\right)  r_t^{(0)z}}{\pi \Gamma\left(\frac{1}{4}\right) K_1}
\ee
where
\be
\label{00}
K_1 = \frac{z^2}{(z+3)(z^2-4)}\frac{\Gamma\left(\frac{1+z}{4}\right)}{\Gamma\left(\frac{3+z}{4}\right)} - \frac{2}{(z-2)(4+z+\beta_z)}
\frac{\Gamma\left(\frac{z+\beta_z}{8}\right)}{\Gamma\left(\frac{z+4+\beta_z}{8}\right)}~.
\ee
Using the expressions for $c_1$ and $c_2$ from eq.\eqref{emzneq2inv} and using eq.\eqref{tempzneq2}, we can write eq.\eqref{delczneq2} as
\be\label{delcompzneq2}
\Delta \mathcal{C}_A =\frac{1}{T_{ent}} \left[ D_1 \Delta E +D_2 \Delta \mu Q\right]
\ee
where $D_1$ and $D_2$ are given by
\bea\label{consd}
D_1&=& \frac{2}{\pi K_1}\left[ \frac{\Gamma \left(\frac{z+3}{4}\right)}{4(z^2 -4)\Gamma \left(\frac{z+5}{4}\right)} 
               -\frac{2}{(z-2)(z+\b_{z} +2)(z+\b_{z}-2)} \frac{\Gamma \left(\frac{z+\b_{z}+4}{8}\right)}
               {\Gamma \left(\frac{z+\b_{z}+8}{8}\right)}\right] \nn
D_2 &=& \frac{1}{\pi z (z+\b_{z} +2)(z+\b_{z}-2)K_1} \frac{\Gamma \left(\frac{z+\b_{z}+4}{8}\right)}{\Gamma \left(\frac{z+\b_{z}+8}{8}\right)}
\eea
and we have used the fact that $Q= m^2 \alpha L \ell=\sqrt{8z(z-1)}$ is the total charge.

\noindent Now the first law of entanglement thermodynamics for the Lifshitz system with $z\neq2$ reads \cite{skarar}
\be\label{thermozneq2}
\Delta E = T_{\rm ent} \Delta S + \frac{K_2}{K_1} \Delta P_x V - \frac{K_3}{K_1} \Delta \mu Q
\ee
where 
\be
\label{01}
K_2 = \frac{\Gamma\left(\frac{1+z}{4}\right)}{(z+2)(z+3)\Gamma\left(\frac{3+z}{4}\right)}~, \quad
K_3 = \frac{\Gamma\left(\frac{z+\b_z}{8}\right)}{2 z (4+z+\beta_z)\Gamma\left(\frac{z+4+\b_z}{8}\right)}~.
\ee
To relate the change in HC with the change in EE, we once again we rewrite eq.\eqref{delcompzneq2} as
\be\label{dezneq2}
\Delta E = \frac{1}{D_1} \Delta \mathcal{C}_A T_{ent} -\frac{D_2}{D_1}\Delta \mu Q
\ee
and equate with eq.\eqref{thermozneq2} to get
\be\label{entcomzneq2}
\Delta \mathcal{C}_A =D_1 \Delta S + \frac{K_2 D_1}{K_1}\frac{\Delta P_x \;V}{T_{ent}} 
				  + \left( D_2 -\frac{K_3 D_1}{K_1} \right) \frac{\Delta \mu\; Q}{T_{ent}}~.
\ee


\section{Conclusions} 

In this paper, the holographic complexity of a $3+1$-dimensional Lifshitz spacetime has been computed. The same has been computed for the 
perturbed Lifshitz spacetime by considering subsystems with the entangling region sufficiently small in size.
This has led to the result for the change in holographic complexity between the excited state and the ground state of Lifshitz spacetime. The analysis has been carried out for both the values of the dynamical scaling
exponent $z$ appearing in the Lifshitz spacetime. The change in the holographic complexity is then expressed in terms of the components
of the holographic stress tensor which in turn is related to the change in the energy and the entanglement chemical potential.
It is observed that the change in pressure does not arise in the expression for the change in the holographic complexity in contrast to the change in entanglement entropy.

\vspace{.5cm}

\section*{Acknowledgements} S.G. acknowledges the support by DST SERB under Start Up 
Research Grant (Young Scientist), File No.YSS/2014/000180.



\begin{thebibliography}{}

\bibitem{Bombelli:1986rw} 
  L.~Bombelli, R.~K.~Koul, J.~Lee and R.~D.~Sorkin,
  ``A Quantum Source of Entropy for Black Holes'',
  Phys.\ Rev.\ D {\bf 34}, 373 (1986).

\bibitem{Srednicki:1993im} 
  M.~Srednicki,
  ``Entropy and area'',
  Phys.\ Rev.\ Lett.\  {\bf 71}, 666 (1993).

\bibitem{Holzhey:1994we} 
  C.~Holzhey, F.~Larsen and F.~Wilczek,
  ``Geometric and renormalized entropy in conformal field theory'',
  Nucl.\ Phys.\ B {\bf 424}, 443 (1994).

\bibitem{Calabrese:2004eu} 
  P.~Calabrese and J.~L.~Cardy,
  ``Entanglement entropy and quantum field theory'',
  J.\ Stat.\ Mech.\  {\bf 0406}, P06002 (2004).
  
\bibitem{Calabrese:2005zw} 
  P.~Calabrese and J.~L.~Cardy,
  ``Entanglement entropy and quantum field theory: A Non-technical 
introduction'',
  Int.\ J.\ Quant.\ Inf.\  {\bf 4}, 429 (2006).
  
\bibitem{Calabrese:2009qy} 
  P.~Calabrese and J.~Cardy,
  ``Entanglement entropy and conformal field theory'',
  J.\ Phys.\ A A {\bf 42}, 504005 (2009).

\bibitem{Eisert:2008ur} 
  J.~Eisert, M.~Cramer and M.~B.~Plenio,
  ``Area laws for the entanglement entropy - a review'',
  Rev.\ Mod.\ Phys.\  {\bf 82}, 277 (2010).

\bibitem{Ryu:2006bv} 
  S.~Ryu and T.~Takayanagi,
  ``Holographic derivation of entanglement entropy from AdS/CFT'',
  Phys.\ Rev.\ Lett.\  {\bf 96}, 181602 (2006).

\bibitem{Ryu:2006ef} 
  S.~Ryu and T.~Takayanagi,
  ``Aspects of Holographic Entanglement Entropy'',
  JHEP {\bf 0608}, 045 (2006).


   \bibitem{Bhattacharya:2012mi}
     J.~Bhattacharya, M.~Nozaki, T.~Takayanagi and T.~Ugajin,
     ``Thermodynamical Property of Entanglement Entropy for Excited States'',
     Phys.\  Rev.\  Lett.\  110, {\bf 091602} (2013).
     
     
           \bibitem{Allahbakhshi:2013rda}
             D.~Allahbakhshi, M.~Alishahiha and A.~Naseh,
             ``Entanglement Thermodynamics'',
             JHEP {\bf 1308} (2013) 102.

\bibitem{skarar}S.~Chakraborty, P.~Dey, S.~Karar, S.~Roy, ``Entanglement thermodynamics for an excited state of Lifshitz system",
JHEP 04 (2015) 133.   
     
     \bibitem{sus1} L. ~Susskind, ``Computational Complexity and Black Hole Horizons", Fortsch.Phys. 64 (2016) 24-43, Addendum: Fortsch.Phys. 64 (2016) 44-48.
     
     
     \bibitem{sus2}  D.~Stanford, L.~Susskind, ``Complexity and Shock Wave Geometries", Phys. Rev. D 90 (2014) 126007.
     
     \bibitem{Mohsen} M.~Alishahiha, ``Holographic complexity",
     Phys. Rev. D 92 (2015) 126009. 
     

   
             
      \bibitem{Ross:2009ar}
        S.~F.~Ross and O.~Saremi,
        ``Holographic stress tensor for non-relativistic theories'',
        JHEP {\bf 0909} (2009) 009.

\bibitem{Kachru:2008yh} 
  S.~Kachru, X.~Liu and M.~Mulligan,
  Phys.\ Rev.\ D {\bf 78}, 106005 (2008).

 \bibitem{Taylor:2008tg} 
   M.~Taylor,
   ``Non-relativistic holography'',arXiv:0812.0530 [hep-th].
 
 
    \bibitem{Ross:2011gu}
      S.~F.~Ross,
      Class.\ Quant.\ Grav.\  {\bf 28} (2011) 215019.
      
     \bibitem{Andrade:2013wsa}
       T.~Andrade and S.~F.~Ross,
       ``Boundary conditions for metric fluctuations in Lifshitz'',
       Class.\ Quant.\ Grav.\  {\bf 30} (2013) 195017.
         
 

    

   
   
   



   
        
 \end{thebibliography}
\end{document}